\RequirePackage[2020-02-02]{latexrelease}
\documentclass[preprint,aps,tightenlines,showkeys,nofootinbib,superscriptaddress]{revtex4}
\usepackage{latexsym}
\usepackage[dvips]{color}
\usepackage{graphicx}
\newcommand{\beq}{\begin{eqnarray}}
\newcommand{\eeq}{\end{eqnarray}}
\newcommand{\bea}{\begin{eqnarray*}}
\newcommand{\eea}{\end{eqnarray*}}
\newcommand{\eq}{eqnarray}

\newcommand{\al}{{\alpha}}
\newcommand{\be}{{\beta}}
\newcommand{\ci}{\cite}
\newcommand{\ga}{{\gamma}}

\newcommand{\ep}{{\epsilon}}

\newcommand{\de}{{\delta}}
\newcommand{\De}{\Delta}
\newcommand{\ka}{\kappa}

\newcommand{\la}{{\lambda}}
\newcommand{\La}{{\Lambda}}
\newcommand{\m}{{\mu}}
\newcommand{\n}{{\nu}}
\newcommand{\si}{{\sigma}}
\newcommand{\Si}{{\Sigma}}
\newcommand{\om}{{\omega}}
\newcommand{\Om}{{\Omega}}
\newcommand{\pa}{{\partial}}
\newcommand{\no}{{\nonumber}}
\newcommand{\f}{\frac}
\newcommand{\ra}{\rightarrow}

\newcommand{\Sch}{Schwarzschild }

\newcommand{\asy}{asymptotically}

\newcommand{\Ho}{Ho\v{r}ava}
\newcommand{\diff}{diffeomorphism}
\newcommand{\DiffF}{${\it Diff}_{\cal F}$}

\begin{document}

\preprint{arXiv:2402.02253v2 [hep-th]}

\title{Rotating Black Holes in
a Viable Lorentz-Violating Gravity:
 Finding
 Exact Solutions Without Tears
}

\author{Deniz O. Devecio\u{g}lu \footnote{E-mail address: dodeve@gmail.com}}

\author{Mu-In Park \footnote{E-mail address: muinpark@gmail.com, Corresponding author}}
\affiliation{ Center for Quantum Spacetime, Sogang University,
Seoul, 121-742, Korea }
\date{\today}

\begin{abstract}
We introduce a two-step procedure for finding Kerr-type rotating black hole solutions without tears.
Considering the low-energy sector of Ho\v{r}ava gravity as a {\it viable} Lorentz-violating gravity in four dimensions
which admits a different speed of gravity, we find the exact rotating black
hole solutions (with or without cosmological constant). We find that the
singular region extends to $r<0$ region from the ring singularity at $r=0$ in Boyer-Lindquist coordinates. There are two Killing horizons
where $g^{rr}=0$ and the black hole thermodynamics laws are still valid.
We find the rotating black hole solutions with electromagnetic charges
only when we consider the {\it noble} electromagnetic couplings, in
such a way that the speed of light is the same as the speed of gravity.
With the noble choice of couplings,
our Lorentz-violating gravity can be consistent with the recently-observed time
delay of the coincident GW and GRB signals. Furthermore, in
Appendices, we show that (a) the uniqueness of the invariant line element $ds^2$ under {{\it the foliation-preserving} diffeomorphism} ${\it Diff}_{\cal F}$, contrary to
{Lorentz-violating}
action, (b) the solutions are the Petrov type I with four distinct principal null vectors, and (c) the Hamilton-Jacobi equation for the geodesic particles are {\it not} separable.

\bigskip
\vspace{1cm}
\noindent
    \emph
   \it{Dedicated to the memory of Roman Jackiw  (8 November 1939 - 14 June 2023).
   }

\end{abstract}


\maketitle

\newpage

\section{Introduction}

The rotating black hole solution which was discovered by Kerr \cite{Kerr:1963} in
general relativity (GR) has occupied an immovable position: The solution structure
is quite rigid and it is extremely difficult to find another {\it exact} solution with non-trivial deformations by cracking the rigid solution structure. Moreover, even deriving the known Kerr solution is not quite easy task and requires some sophisticated methods whose generic validities in other gravity theories are not clear \cite{Chan:1985}.
In particular, from the separability of Hamilton-Jacobi equation for the geodesic particles in a quite general context, Carter obtained the Carter constant as well as the Kerr solution \cite{Cart:1973}. Later, it was proved that the Petrov type-D solutions, like Kerr metric, guarantee the existence of a Killing tensor and its concrete form was derived in GR \cite{Walk:1970}

In this paper, we introduce a two-step procedure to find Kerr-type exact solutions without much tears. We apply the procedure
to the low-energy sector of Ho\v{r}ava gravity as a Lorentz-violating (LV)
gravity in four dimensions which admits a different speed
of gravity, and
we find the exact rotating black hole solutions (with or without cosmological
constant). In addition, if we consider charged rotating black holes,
the {\it noble} electromagnetic couplings are needed for consistency, in such
a way that the speed of light is the same as the speed of gravity. In Sec. II,
we introduce our two-step procedure to find Kerr-type exact solutions.
In Sec. III, we apply the procedure to the low-energy sector of Ho\v{r}ava
gravity in four dimensions and find the exact rotating black hole solution
without cosmological constant. In Sec. IV, we study the singularity structure
which is richer than Kerr solution. In Sec. V, we study the horizon structure
and new barriers for geodesic particles which are genuine to our LV gravity.
In Sec. VI, we consider the generalizations with cosmological constant and
electromagnetic charges, and
show that the first law of black hole thermodynamics is satisfied. In Sec. VII, we consider some observational constraints and in Sec. VIII, we conclude with some discussions. In Appendix {\bf C}, we prove the uniqueness of the invariant line element $ds^2$ under {{\it the foliation-preserving} diffeomorphism} ${\it Diff}_{\cal F}$, contrary to widespread belief based on LV action.
In Appendix {\bf D}, we study the Petrov classification,
the Killing tensor, and the Hamilton-Jacobi equation.

\section{A two-step procedure to find Kerr-type stationary solutions}

(i) Suppose that there is an {\it exact} ``massless" rotating space-time solution, which means it is the exact solution for an {\it arbitrary} rotation parameter $a$. This is the first step and the massless solution is a {\it seed} solution for the step 2.

(ii) The second step is to introduce the ``mass"-dependent ansatz functions into the {\it massless} seed solution,
without loss of much generality but still in a solvable way, {\it i.e.}, with
ordinary differential equations {(ODE)} for the polar angle $\theta$ or the radial coordinate $r$. This requires noble insights as well as some trials and errors. Without cosmological constant, we find that the situation becomes simpler since the
{\it massive} rotating solution does not deform the angle-dependent parts but just
the radial-dependent parts
of the {\it massless} rotating space-time solution. On the other hand, with a cosmological constant, it is a bit more complicated and we need a more clever choice of the ansatz. So, in the following sections, we first consider the solution without cosmological constant and later with a cosmological constant.

The above two-step procedure can be applied to any gravity theory, in principle. In this paper, we apply the procedure to find the rotating black hole solutions in the low-energy sector of Ho\v{r}ava gravity in four dimensions, for the first time. {The second step looks a quite vague procedure due to
the absence of a systematic way to find the viable ansatz. But, compared to
other usual approaches, the viable ansatz space is much reduced such that the
original (partial-differential)
equations of motion reduced to ODE, which has not been succeed before. Actually,
this corresponds to Kerr-Schild approach in GR \cite{Kerr:1965} to write the
Kerr metric \cite{Kerr:1963} as the massless rotating solution, perturbed by
its mass. The key point of this second-step procedure is to find the viable ODE,
by separating the angle-dependent rotating metric, parameterized by the spin $a$,
and the additional ($r$-dependent,
without cosmological constant)
corrections due to the
mass $m$. Of course, this does not always mean the existence of the exact solutions which crucially depends on the explicit form of an ansatz and the structure of ODE.}

\section{Low-energy sector of Ho\v{r}ava gravity and rotating black hole solutions }

The low energy sector of (non-projectable) Ho\v{r}ava gravity \cite{DeWi:1967,Hora:2009} is described by the action, up to boundary terms,
\begin{\eq}
S_g &= & \int_{{\bf R} \times \Si_t} dt d^3 x
\sqrt{g}N\left[\frac{1}{\kappa}\left(K_{ij}K^{ij}-\lambda
K^2\right)+\xi R^{}-2 \La+\f{\sigma}{2} a_i a^i \right]\ ,
\label{horava}
\end{\eq}
where
$
 K_{ij}=({2N})^{-1}\left(\dot{g}_{ij}-\nabla_i
N_j-\nabla_jN_i\right)
$
is the extrinsic curvature [the overdot $(\dot{})$ denotes the time derivative and $\nabla_i$ is the covariant derivatives for the induced metric $g_{ij}$ on the time-slicing hypersurface $\Si_t$] and $R$ is the {\it three}-curvature in the ADM metric
\begin{\eq}
\label{metric}
ds^2=-N^2 c^2 dt^2+g_{ij}\left(dx^i+N^i dt\right)\left(dx^j+N^j
dt\right),
\end{\eq}
respectively. (Hereafter, we shall use the unit $c=1$, unless stated otherwise.)
The last term in the gravity action (\ref{horava}) is introduced for completeness, with the proper acceleration $a_i=\nabla_i ln N$ \cite{Blas:2009} \footnote{{In \cite{Blas:2009b}, it was argued that the non-projectable Ho\v{r}ava gravity {\it without} $a_i$ term is problematic due to the strong-coupling and instability problems of an extra scalar graviton mode in a perturbative analysis. However, later it has been proved, in a {\it non-perturbative} analysis, that the true degrees of freedom of the low-energy limit of Ho\v{r}ava gravity, which is crucial to the above problems, is just two as in GR, corresponding to the usual {\it transverse-traceless (TT)} mode, such that the scalar graviton mode and so their involving problems are absent non-perturbatively \cite{Bell:2010,Deve:2020}.
}}.

The equations from the variations of $N, N^i$, and $g^{ij}$ are given by
\beq
{\cal H}&\equiv&\f{1}{\kappa}\left(K_{ij}K^{ij} -\lambda K^2\right)-\xi R +2 \La
-\si \left(\f{1}{2} \f{\nabla_i N \nabla^i N }{N^2} -\f{\nabla_k \nabla^k N }{N} \right) =0\, , \label{eom1} \\
{\cal H}^i&\equiv&{-}\f{2}{\ka}\nabla_j \left(K^{ji}-\lambda\,Kg^{ji}\right)=0\, ,\label{eom2}\\
E_{ij}&\equiv&\frac{1}{\kappa}\left( E_{ij}^{(1)}-\lambda E_{ij}^{(2)} \right)
+\xi E_{ij}^{(3)}
+\frac{\si}{2}E_{ij}^{(4)}=0, \label{eom3}
\eeq
where
\bea
E_{ij}^{(1)}&=& N_i \nabla_k K^k{}_j + N_j\nabla_k K^k{}_i -K^k{}_i
\nabla_j N_k-
   K^k{}_j\nabla_i N_k - N^k\nabla_k K_{ij}\no\\
&& - 2N K_{ik} K_j{}^k
  -\frac{1}{2} N K^{k\ell} K_{k\ell}\, g_{ij} + N K K_{ij} + \dot K_{ij}
\,,\no \\
E_{ij}^{(2)}&=& \frac{1}{2} NK^2 g_{ij}+ N_i \pa_j K+
N_j \pa_i K- N^k (\pa_k K)g_{ij}+  \dot K\, g_{ij}\,,\no\\
E_{ij}^{(3)}&=&N\Big(R_{ij}- \frac{1}{2}R g_{ij}+\frac{\La}{\xi}g_{ij}\Big)-(
\nabla_i\nabla_j-g_{ij}\nabla_k\nabla^k)N\,,\no\\
E_{ij}^{(4)}&=&\f{1}{N} \Big(-\f{1}{2} g_{ij} \nabla_k N \nabla^k N+ \nabla_i N \nabla_j N\Big)\,.
\eea
With the arbitrary parameters $\la, \xi$, and $\si$, the action does not admit the full diffeomorphism but only the {\it foliation-preserving} \diff~(${\it Diff}_{\cal F}$) \cite{Hora:2009,Park:2009},
\begin{\eq}
\delta_{\xi} t&=&-{\xi}^{0}(t),~~ \delta_{\xi} x^{i}=-\xi^{i}(t,{\bf x}), \no
\\
\delta_{\xi} N&=&(N{\xi}^{0})_{,0}+\xi^{k}\nabla_{k}N, \no 
\\
\delta_{\xi}{N_{i}}&=&{\xi}^{0}{}_{,0}N_{i}+\xi^{j}{}_{,0}g_{ij}
+\nabla_{i}\xi^{j}N_{j}
+N_{i,0}{\xi}^{0}+\nabla_{j}N_{i}\,\xi^{j}, \no 
\\
\delta_{\xi}{g_{ij}}&=&\nabla_{i}\xi^{k}g_{kj}+\nabla_{j}\xi^{k}g_{ki}
+g_{ij,0}{\xi}^{0}.
\label{delg3}
\end{\eq}
Note that the Einstein-aether theory reduces to the same gravity action (\ref{horava}) with the {\it hypersurface-orthogonal} aether field \cite{Jaco:2013} or the khronometric theory with a {\it global} time-like khronon field \cite{Blas:2011}. In this paper, we do not introduce additional assumptions on the aether or khronon field, but we only study the gravity action (\ref{horava}).

The first step toward the rotating black hole solutions is to find the {\it massless} rotating black hole solutions. Here, we first consider the
asymptotically {\it flat} case without cosmological constant and the $a_i$
extension, {\it i.e.}, $\La=0,~ \si=0$, for simplicity. To this end, we note
the recent finding by one of authors that the massless Kerr solution is also
a solution of Ho\v{r}ava gravity with an arbitrary $\la$ \cite{Park:2023},
\beq
ds^2_{0}=-\frac{\rho^2 {\Delta}_r^{(0)} }{ {\Sigma}^2_{(0)}} dt^2+\frac{{\rho}^2}{ {\Delta}_r^{(0)}}dr^2+ {\rho}^2 d \theta^2
+\frac{{\Sigma}^2_{(0)} \mbox{sin}^2\theta}{{\rho}^2} d \phi ^2,
\label{massless_Kerr}
\eeq
where,
$
{\rho}^2 = r^2 + a^2 \mbox{cos}^2 \theta,
{\Delta}_r^{(0)} = \left( r^2 + a^2 \right) ,
{\Sigma}^2_{(0)} = \left( r^2 + a^2 \right) {\rho}^2,
$
and $(t,r,\theta, \phi)$ are the Boyer-Lindquist coordinates
\ci{Cart:1973,Gibb:1977,Park:2001}. The metric (\ref{massless_Kerr})
is nothing but the {\it flat} Minkowski spacetime
written in the {\it ellipsoidal} spacial
coordinates, which are related to Cartesian coordinates
$x=(r^2+a^2)^{1/2} \mbox{sin}\theta~ \mbox{cos} \phi,
y=(r^2+a^2)^{1/2} \mbox{sin} \theta~ \mbox{sin} \phi,
z=r~\mbox{cos} \theta$.


Here, it is important to note that $r=0$
is not the end of the coordinates but there is another copy of Kerr spacetime
in the $r<0$ regime, with another asymptotic infinity at $r\ra -\infty$ in ellipsoidal coordinates system. Actually, the massless Kerr metric (\ref{massless_Kerr}) was interpreted as a wormhole solution in \cite{Gibb:2017}. Of course, there is no event horizon in (\ref{massless_Kerr}) but {\it the genuine spacetime deformation for a rotating object is believed to be naturally encoded in the ellipsoidal coordinates with the rotation parameter $a$.}

The next step towards the generic rotating black hole solutions is to consider the mass-dependent ansatz functions without loss of generality, {\it until we may obtain the consistent solutions}. In the known Kerr metric, the mass term is tightly bounded to the rotation parameter $a$ and it is not easy to separate them. So, it is important to consider a generic ansatz such that the spin part and the mass part are consistently separated which might crack the rigid structure of Kerr solution. Now, by comparing to Kerr solution, we consider the metric ansatz,
\beq
ds^2_{1}=-N^2 dt^2+\frac{\rho^2}{ \Delta_r}dr^2+ \rho^2 d \theta^2
+\frac{ \Sigma^2 \mbox{sin}^2\theta}{\rho^2} \left(d \phi +N^{\phi} dt \right)^2,
\label{massive_Kerr_metric}
\eeq
where we introduce
\beq
\Sigma^2&=& \left( r^2 + a^2 \right) \rho^2 +f(r) a^2 \mbox{sin}^2\theta, \no \\
N^2&=&\frac{\rho^2 \Delta_r (r) }{ \Sigma^2},~~ 
N^{\phi}=-\f{g(r)}{\Sigma^2}
\label{massive_Kerr_functions}
\eeq
with three $r$-dependent undetermined functions $f(r), g(r)$, and $\Delta_r (r)$. Then, by solving the full equations of motion (\ref{eom1}-\ref{eom3}) for the ansatz (\ref{massive_Kerr_metric}) and (\ref{massive_Kerr_functions}), we can determine the undetermined functions uniquely as,
\beq
f(r)=2m r, ~g(r)=2 {\it a} mr \sqrt{\kappa \xi}, ~\Delta_r (r)=r^2+a^2-2mr,
\label{f,g_sol}
\eeq
which reduce to the known Kerr solution in the GR case of $\xi=1/\ka$ or Schwarzschild solution with $a=0$, where $m$ is an integration constant parameter. Here, we adopt the usual choice of $N^{\phi}|_{\infty}=0, W(\infty) \equiv N \sqrt{g_{rr}}|_{\infty}=1$. It is rather surprising that the Kerr-solution cracking term with the LV factor $\sqrt{\kappa \xi}$ appears only in $N^{\phi}$. But, if we look at the component form
\beq
ds^2_{1}&=&\left[ -\f{(\De_r-a^2 \mbox{sin}^2\theta )}{\rho^2}+\f{(\ka \xi-1)~(2mr)^2~ a^2 \mbox{sin}^2\theta }{\rho^2 \Si^2}\right] dt^2 \no \\
&+&\frac{\rho^2}{ \Delta_r}dr^2+ \rho^2 d \theta^2
+\frac{ \Sigma^2 \mbox{sin}^2\theta}{\rho^2} d \phi ^2-\f{4 {\it a} mr \sqrt{\ka \xi}~ \mbox{sin}^2\theta}{ \rho^2} dt d\phi,
\label{massive_Kerr_metric_comp}
\eeq
one can easily see the non-trivial LV
effect for $\xi \neq 1/\ka$
in $g_{tt}$ as well as in $g_{t \phi}$ components. Another notable property is that our
solution, as well as the Kerr solution with $\xi = 1/\ka$, are valid for
an {\it arbitrary} $\la$ due to $K=0$, {\it i.e.,} ``maximal"
slicing. This makes even Kerr solution or Schwarzschild solution with $a=0$ has different notions of singularities as we can see in the next section, due to the lack of the full {\it Diff} with $\la \neq 1$.

The mass and angular momentum are computed as
\beq
M=16 \pi \xi m, ~ J=16 \pi {\it a} m \sqrt{ \xi \ka^{-1}},
\label{M,J}
\eeq
by the standard definition \cite{Bana:1992}, as the conjugates to $W(\infty)$ and $N^{\phi} (\infty)$, respectively, in the boundary term
\beq
{\cal B}=(t_2-t_1) [-W(\infty) M+N^{\phi} (\infty) J ]
\label{B}
\eeq
which makes the total action $S=S_g +{\cal B}$ to be {\it differentiable} with the boundary
conditions $\de N^{\phi} (\infty)=\de W(\infty)=0$. We obtain the same result in the
recently-proposed conserved-charge formula in a {\it covariant} formalism
\cite{Deve:2021}
and it provides another non-trivial evidence of the formalism
{(see Appendix {\bf A} for the details)}. The generalizations
to the asymptotically {\it (A)dS} black holes ($\La \neq 0$) and the {\it charged} black holes will be
considered in the later part, due to some complications. However, with the $a_i$
extension term, we find
no solution for the ansatz (\ref{massive_Kerr_metric}) and (\ref{massive_Kerr_functions}), except the massless solution, which indicates that, with the $a_i$ term, the solution structure will be quite different than ours or GR. So, from now on we consider only the standard case without $a_i$, {\it i.e.,} $\si=0$.

\section{Singularity structure}

Due to the {\it apparent} lack of the full {\it Diff}, we have a different notion of singularities. In our case, the physical singularities are captured by \DiffF~ invariant
curvatures
($K=0,~K_{ij} R^{ij}=0$),
\beq
R \sim \f{a^2 m^2}{\rho^6 \Si^4},~R_{ij} R^{ij} \sim \f{m^2}{\rho^{12} \Si^8},~K_{ij} K^{ij} \sim \f{\ka \xi a^2 m^2}{\rho^6 \Si^4},
\label{3D_curvature}
\eeq
where we have omitted the other detailed factors which are finite and
are not canceled by the denominators. It is interesting to note that the
omitted factors are not modified by our LV solutions and they are exactly the
same as in GR. The distinct feature in our case is that the singularity
properties of each quantity in (\ref{3D_curvature}) are physically meaningful
and they show the curvature singularities at $\Si^2=0$, as well as the usual
singularity at $\rho^2=0$. On the other hand, the four-dimensional curvature invariants in GR are given by
\beq
&&R^{(4)} \sim (\ka \xi-1)~ \f{a^2 m^2}{\rho^6 \Si^4},~R^{(4)}_{\m \n} R^{(4) \m \n} \sim (\ka \xi-1)^2~ \f{a^4 m^4}{\rho^{12} \Si^8},\no \\
&&R^{(4)}_{\m \n \si \rho} R^{(4) \m \n \si \rho} \sim (\ka \xi-1)~ \f{m^2}{\rho^{12} \Si^8}+\f{m^2}{\rho^{12}} \left(\cdots\right),
\label{4D_curvature}
\eeq
where the omitted term in $\left(\cdots\right)$ is the same as in GR. This clearly shows that the additional singularities of (\ref{3D_curvature}) at $\Si^2=0$ are exactly canceled in the GR limit $\xi =\ka^{-1}$ so that only the usual
ring singularity at $\rho=0$, {\it i.e.,} $r=0, \theta=\pi/2$ remains in GR \footnote{This is a common feature in \Ho~gravity \cite{Lu:2009,Park:2012}, as was once noted by S. Mukohyama earlier (a private communication). }. Furthermore, even in the GR limit $\xi =\ka^{-1}$, {\it i.e.}, Kerr solution or even \Sch solution, due to the physical meaningfulness of \DiffF~ invariant quantities in (\ref{3D_curvature}), not (\ref{4D_curvature}), we still have the same additional singularities at $\Si^2=0$.

Fig. 1 shows the singularity surfaces of
\beq
\mbox{cos}^2\theta=\f{2mr a^2 +r^2 (r^2+a^2)}{a^2 (2 mr -(r^2+a^2))}
\eeq
for $\Si^2=0$, by varying the mass parameter $m$ and the rotation parameter $a$. The lower $(r<0)$ and the upper $(r>0)$ branches correspond to $m>0$ and $m<0$, respectively (the left panel).

As $m \ra \pm \infty$, the singularity surfaces cover the whole region of $r<0$ or
$r>0$, with $r_0=\mp a/\sqrt{3}$ which touches the $\theta$ boundaries
$\theta=0, \pi$
(the right panel).
As $a \ra 0$, the singularity surfaces reduce to the point singularity of \Sch black hole at $r=0$. Other than these extreme cases, {\it i.e.,} $0<a \leq m <\infty$, there exits always a time-like trajectory which avoids the singularity surfaces such that the closed-time-like (CTC) curves  are possible, similarly to Kerr solution in GR.

\begin{figure}
\includegraphics[width=7cm,keepaspectratio]{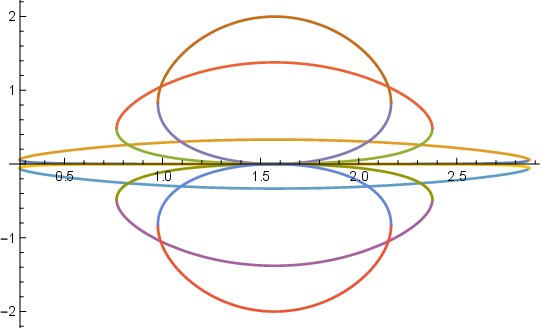}
\qquad
\includegraphics[width=7cm,keepaspectratio]{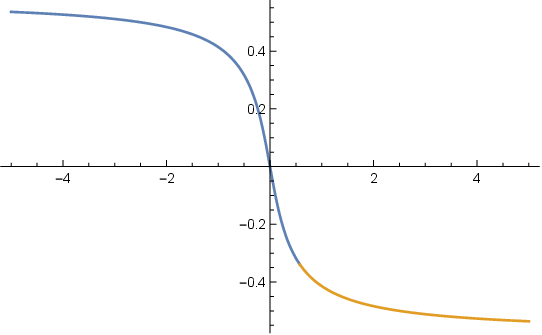}
\caption{Left: $r$ vs. $\theta~ [0, \pi]$ of singularity surfaces for $\Si^2 (r, \theta)=0$, by varying $a=2,1,0.1$ (from the outer to inner curves) with $m=2$ (the lower branch), $m=-2$ (the upper branch). Right: The extreme radius $r_0$ vs. $m$, where $\theta$ becomes an extreme value for a given $a$ and $m$, and
$r_0
\ra \mp a/\sqrt{3}$ as $m \ra \pm \infty$.}\label{fig:singular}
\end{figure}

\section{Killing horizons and particle barriers}

In our coordinates, there are two event horizons at $r_{\pm}$ where $g^{rr}=\Delta_r/\rho^2=0$, which occurs when
\beq
\De_r =r^2+a^2-2ma=0,
\eeq
since $\rho^2 >0$, excluding the singularity at $\rho^2=0$ that was studied in Sec. IV. The event horizons at $r_{\pm}$ are Killing horizons with the Killing vectors $\chi^a_{\pm}=(1,0,0,\Om_{\pm})$, which become {\it null} vectors $\chi^a_{\pm} {\chi_a}_{\pm}=-\rho^2 \De_r/\Si^2+\Si^2 \mbox{sin}^2\theta (N^{\phi}-\Om_{\pm})^2/\rho^2=0$ for the {\it constant} angular velocities on the horizons $r_{\pm}$, $\Om_{\pm}=-N^{\phi}|_{r_{\pm}}$. Then, one can define the surface gravities $\ka_{\pm}$ such that $\chi^{\nu}_{\pm} \nabla_{\nu}^{(4)} \chi^{\mu}_{\pm}=\ka_{\pm} \chi^{\mu}_{\pm}$ {\it on} the horizons $r_{\pm}$ \footnote{The direct computation in our coordinates is rather tricky and we first need to consider $\chi^{\nu}_{\pm} \nabla_{\nu}^{(4)} \widehat{\chi}^{\mu}_{\pm}=\ka_{\pm} \widehat{\chi}^{\mu}_{\pm}$ with $\widehat{\chi}^{\mu}_{\pm}=(1,\f{\De_r}{r^2+a^2},0,\Om_{\pm})$ and then consider the limit $\widehat{\chi}^{\mu}_{\pm} \ra \chi^{\mu}_{\pm}$ at the horizons.}.

From the {\it hypersurface-orthogonality} of the Killing horizons, one can find that $\ka_{\pm}$ are constants on the corresponding horizons $r_{\pm}$, {\it i.e.},
\beq
\chi^{[ \m} \nabla^{\n ]}_{(4)} \ka_{\pm}=-\chi^{[ \m} {R^{\n ]}}_{ \si (4)} \chi^{\si}= -\chi^{[ \m} {T^{\n ]}}_{\si ({\it eff}) } \chi^{\si}=0,
\label{0th law}
\eeq
where we have expressed our non-covariant equations (\ref{eom1}-\ref{eom3}) into
the {\it covariant} form
$R_{\m \n}^{(4)} -(1/2) g_{\m \n}^{(4)} R^{(4)}=T_{\m \n}^{({\it eff})}$ with
an {\it effective} energy-momentum tensor,
\beq
{{T}^{\m}}_{\n (\it{eff})}=\left( \begin{array}{cccc}
\widehat{\rho} & 0 & 0 & 0 \\
0 & \widehat{p_1} & \widehat{p_2} &0 \\
0 & \widehat{p_3} & -\widehat{p_1} & 0 \\
\widehat{p_4} & 0 & 0 & -3 \widehat{\rho},
\label{T_eff}
\end{array}\right)
\eeq
which is a {\it non-perfect} fluid form
(see Appendix {\bf B} for the explicit expressions of $\widehat{\rho}$ and $\widehat{p_a}$),
but still satisfies the covariant conservation equation $\nabla_{\m}^{(4)} T^{\m \n}_{({\it eff})}=0$.
The effective energy-momentum violates the dominant energy condition,
${T^{0}}_{ 0 (\it{eff})}\geq | {T^{\m}}_{\n (\it{eff})}|$, especially by
$|{T^{3}}_{3 (\it{eff})}|=3 |\widehat{\rho}|>\widehat{\rho}$, but it satisfies
$\chi^{[ \m}{T^{\n ]}}_{\si (\it{eff})} \chi^{\si}=0$
on the horizons $r_{\pm}$ so that the {\it zeroth} law (\ref{0th law}) is satisfied. In other words, the usual notion of Killing horizons at $r_{\pm}$ satisfying the zeroth law are still valid in our LV gravity, which seems to be a quite non-trivial result since the effective energy-momentum tensor $T_{\m \n}^{(\it{eff})}$ might break the last equation of (\ref{0th law}), generally.

Now, we turn to the discussion on the role of the event horizons to the geodesic particles to see whether they have the similar roles even in our LV gravity also. To this end, we consider the conserved energy $E=-v^{\m} (\partial_t)_\m$ and angular momentum $L=v^{\m} (\partial_{\phi})_\m$, for the time-like Killing vector $(\partial_t)_\m$ and the axial Killing vector $(\partial_{\phi})_\m$, and the {\it on-shell} equations
\beq
-\ep =g_{\m \n} v^{\m} v^{\n},
\label{geodesic}
\eeq
where $v^{\m} ={x^{\m}}'
\equiv d x^{\m}/ d \tau$ for a particle's $4$-velocity with the proper time $\tau$, and $\ep=+1~ (-1)$ for a time-like (space-like) geodesics and $\ep=0$ for a null geodesics.

Here, it is important to note that {\it the line element (\ref{metric}) describes
the invariant distance measure even in our LV case}, contrary to a widespread
belief based on the symmetry transformation of action which allows 
LV action (\ref{horava}) with
the {\it reduced} \diff~\DiffF. Actually, one can prove that the {\it invariant} line element is only given by  (\ref{metric}), due to mixing between different components under the coordinate transformation, even if we consider
\DiffF~(see Appendix {\bf C} for the proof). This justifies $E$ and $L$ as
the conserved quantities, {\it independently on the chosen coordinates}, even
in \Ho~ gravity,
which has {\it never} been {\it known} in the literature.

Then, by solving ${t'}, {\phi'}$, and ${r'}$ in terms of the conserved quantities $E$ and $L$, one obtains
\beq
{t'}&=&\f{\left[(r^2+a^2) \rho^2+2mra^2 \mbox{sin}^2 \theta\right] E-2mra \sqrt{\ka \xi} L}{\rho^2 \De_r},\\
{\phi'}&=& \f{2mra \sqrt{\ka \xi} E+g_{tt} \rho^2 L/\mbox{sin}^2 \theta}{\rho^2 \De_r},\\
{r'}&=&\pm \sqrt{-2 \left(V+g_{\theta \theta} {\theta'}^2/2 g_{rr} \right)},
\eeq
where
\beq
V&=&-\f{\ep mr}{\rho^2}+\f{L^2}{2 \mbox{sin}^2\theta \rho^2}+(\ep-E^2) \f{(r^2+a^2)}{2 \rho^2}-\f{mr}{\rho^4} \left(-a E \mbox{sin}\theta+\f{L}{\mbox{sin}\theta} \right)^2 \no \\
&&+\f{ 2 (\sqrt{\ka \xi}-1) mra  E L}{\rho^4}- \f{(\ka \xi-1)  (2 mra)^2 L^2}{2 \rho^4 \Si^2}.
\eeq
For a complete analytic integration of the geodesics, including $\theta$ direction, we would need another constant of motion (we will discuss more about this in Appendix {\bf D}) but there is an important case where we can neglect ${\theta'}$, that is the {\it equatorial} geodesics, $\theta=\pi/2$: For other angles, $\theta \neq \pi/2$, the {\it full} geodesic equation (\ref{geodesic}) tells us that the $\theta=constant$ plane is not maintained generally, due to ${\theta''} \neq 0$ even if ${\theta'}=0$ initially.

Then, for the equatorial and the {\it co-rotating} geodesics with $L=a E$, which corresponds to the radial geodesics in non-rotating geometries \cite{Chan:1985}, we obtain [$+~ (-)$ denotes the out (in)-going geodesics]
\beq
\f{dt}{dr}&=&\pm \f{(r^2+a^2)-2 (\sqrt{\ka \xi}-1) ma^2 r^{-1} }{\De_r \sqrt{-U}},\label{dtdr}\\
\f{d \phi}{ dr} &=& \pm \f{a \left[ 1+2 (\sqrt{\ka \xi}-1) m r^{-1} - (\ka \xi-1) (2ma)^2 \left( (r^2+a^2)r^2+2 mr a^2\right)^{-1}\right]}{\De_r \sqrt{-U}},\label{dpdr}
\eeq
where
\beq
-U=-\f{\ep \De_r}{E^2 r^2}+1-\f{ 4 (\sqrt{\ka \xi}-1) ma^2 }{r^3}
+\f{(\ka \xi-1)  (2 ma^2)^2}{r^2 [(r^2+a^2) r^2 +2mr a^2]}.
\eeq
Note that (\ref{dtdr}) and (\ref{dpdr}) show the characteristic behaviors of the {\it light-cones}, which ``close up" or ``peel off", as the event horizons at $r_{\pm}$ are approached with $\De_r\ra 0$, especially for the null geodesics $(\ep=0)$, {\it independently} on the particle's energy $E=L a^{-1}$.

\begin{figure}
\includegraphics[width=7cm,keepaspectratio]{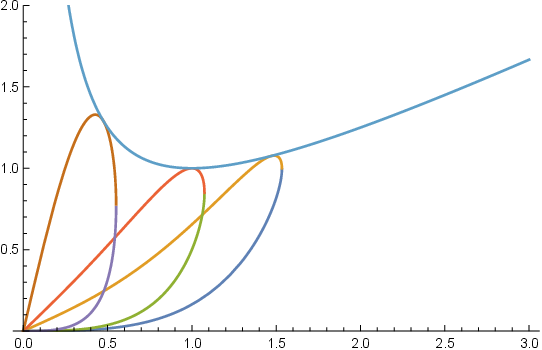}
\qquad
\includegraphics[width=7cm,keepaspectratio]{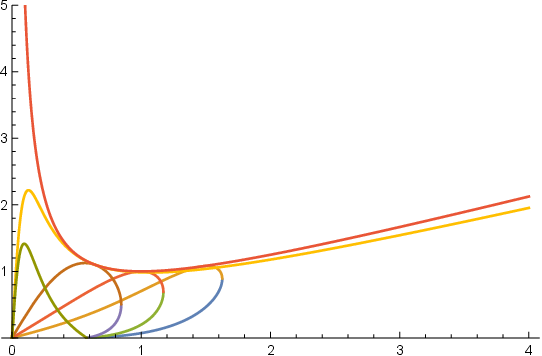}
\caption{Mass parameter $m$ vs. horizons $r_{\pm}$ (the top curves) or barriers (the bottom curves) for null (the left panel) and time-like (the right panel) trajectories. The bottom curves denote the barrier regions by varying $E$ and $\ka\xi$: For the null trajectories (left panel), we take $\ka\xi=1.5,4,10$ (left to right); for the time-like trajectories (right panel), we take $\ka\xi=2, E=0.3$, and $\ka\xi=2,4,10, E=2$ (left to right). For time-like cases (right panel), the orbits are bounded for $E<\ep$ (yellow curve). The barriers exist even when the event horizons $r_{\pm}$ are absent when $m<m_*$, with the extremal black hole mass $m_*$, where two horizons merge.}
\label{fig:barrier}
\end{figure}

Moreover, with the coupling $\ka \xi > 1$ for null particles but arbitrary
$\ka \xi$ for time-like particles, it
shows a potential barrier, {\it i.e.}, $U > 0$. From the condition that the barrier should not exist outside the even horizon $r_+$ so that the identity of black holes can be still maintained, one finds the condition $\ka \xi \leq 4$, which gives its physical upper bound (Fig. 2): In the GR limit of $\ka \xi = 1$, the barrier also exists for time-like cases (the right panel) though absent in null cases (the left panel), but it exists always inside the event horizons and so it is not harmful. It is notable also that, for ingoing geodesics, the barrier can act as a {\it classical} protector of the curvature singularities at $r \leq 0$ with $m>0$. On the other hand,
for outgoing geodesics, it can produce thermal radiations via {\it quantum-mechanical} tunneling from the inner region even when the event horizon is absent.

\section{Generalizations with cosmological constant and electromagnetic charges}

Generalizing our rotating solution to the \asy~ {\it(A)dS} spacetime with a
cosmological constant $\La$ is not quite straightforward and needs a quite
clever choice of the reference solution, as given by \cite{Gibb:2004}
\footnote{We thank {\"O}
Sar{\i}o\u{g}lu for informing \cite{Gibb:2004}, which turns
out to be a good reference solution for {\it Kerr-(A)dS} solutions.}. Starting
from the Kerr-(A)dS solution of \cite{Gibb:2004}, and doing the same two-step
procedure as in the previous sections, with the first step of finding the
massless solution
done also in \cite{Park:2023}, we obtain the rotating {\it (A)dS} black hole solutions as
\beq
ds^2=-N^2 dt^2+\frac{\rho^2}{\Delta_r (r)}dr^2+ \frac{\rho^2}{\Delta_{\theta} (\theta)} d \theta^2
+\frac{\Sigma^2 \mbox{sin}^2\theta}{\rho^2 \Xi^2 }
\left(d\phi+ N^{\phi} dt \right)^2,
\label{KAdS_ansatz}
\eeq
where
\begin{eqnarray}
{\Xi}&=&1+\f{\La a^2 }{ 3 \xi}, ~
{\Delta}_\theta = 1+\f{\La a^2 \mbox{cos}^2 \theta}{ 3 \xi},\nonumber \\
{\Delta}_r &=& \left( r^2 + a^2 \right) \left(1 -\f{\La r^2}{3 \xi}\right)-2mr, \no \\
{\Sigma}^2 &=& \left( r^2 + a^2 \right) \rho^2 {\Xi}+f(r) a^2 \mbox{sin}^2\theta, \no \\
N^2 &=&\frac{\rho^2 \Delta_r \De_\theta}{\Sigma^2},~
N^{\phi}=-\f{g(r) \De_{\theta} }{\Si^2}
\label{KAdS_sol}
\end{eqnarray}
with the same solutions for $f(r)$ and $g(r)$ as in (\ref{f,g_sol}).

The mass and angular momentum can be computed  as {(see Appendix {\bf A} for the details)}
\beq
M=\f{16 \pi \xi m}{\Xi^2}, ~J=\f{16 \pi a m \sqrt{\xi \ka^{-1}}}{\Xi^2},
\label{M,J_KdS}
\eeq
similarly to the $\La=0$ case in Sec. III, and they satisfy the first law of black hole
thermodynamics,
\beq
dM=T_H dS_H +\Om_{H} dJ
\label{1st law}
\eeq
with the horizon entropy,
\beq
S_H=\f{\xi (r_{H}^2+a^2)}{4 \hbar}
\eeq
and the Gibbons-Hawking temperature $T_H$ at the black hole horizons $r_{\pm}$
and the cosmological horizon $r_{++}$
\footnote{Here, we take the convention of the {\it negative} temperatures for
the inner and cosmological horizons
\cite{Curi:1979,Klemm:2004,Park:2006,Cveti:2018}.}, and angular velocity at the horizons $\Om_H$,
\beq
T_H&=&\f{\ka_{H} \hbar}{2 \pi}=\f{r_H \left(1-\f{a \La}{3 \xi}-\f{r_H^2 \La}{\xi}-\f{a^2}{r_H^2} \right)}{4 \pi(r_H^2+a^2)}, \no \\
\Om_{H}&=&-N^{\phi}|_H=\f{a \left(1-\f{r_H^2 \La}{3 \xi} \right) \sqrt{\ka \xi}}{r_H^2+a^2}.
\eeq

Note that $\Om_{\infty}=-N^{\phi}|_{\infty}=0$, contrary to another standard form
of Kerr-(A)dS solution in GR \cite{Cart:1973,Gibb:1977,Hawk:1998,Henn:1985}, and
we obtain the correct $\Om_H$ for the first law of thermodynamics, without the
background subtraction $\Om_H-\Om_{\infty}$. We obtain the correct mass and
angular momentum from the standard definition or from the recently-proposed
charge formula for a covariant formalism, consistently with the first law of
thermodynamics \cite{Cald:1999,Gibb:2004b}. Moreover, due to the
{\it r-independence} of the conserved charge
in \cite{Deve:2021}, we obtain the mass and angular momentum without the conceptual problem of the infinite boundary in the standard method for the rotating {\it dS} solutions.

Another standard form of rotating (A)dS black hole solutions with a non-vanishing angular velocity $\Om_{\infty}$ at infinity \cite{Cart:1973,Gibb:1977,Henn:1985,Hawk:1998,Cald:1999} is obtained by considering
\beq
\widehat{N}^{\phi}&=&-\f{g(r) \De_{\theta} }{\Si^2}-\f{a \La}{3 \xi},
\eeq
via a coordinate transformation $d \phi=d \widehat{\phi}-(a \La/3 \xi)dt$ in
the metric (\ref{KAdS_ansatz}), which
corresponds to a \DiffF \cite{Cart:1973,Henn:1985,Park:2023}.
We find the same mass and angular momentum as (\ref{M,J_KdS}) and the first law of black hole thermodynamics as (\ref{1st law}), but now with the subtracted angular speed $\widehat{\Om}_{H}=-\widehat{N}^{\phi}|_H+\widehat{N}^{\phi}|_{\infty}$, in agreement with GR's result \cite{Hawk:1998,Henn:1985}.

Adding the electromagnetic charges to the rotating black hole solutions by coupling Maxwell fields to \Ho~ gravity is also an important generalization. To this ends, we consider the LV Maxwell action, up to boundary terms,
\beq
S_M=\int_{{\bf R} \times \Si_t} dt d^3 x \sqrt{g} N \left[ -\f{2 \eta}{N^2} \left(E_i +F_{ij}N^j \right)^2+\zeta F_{ij} F^{ij}\right],
\eeq
which admits \DiffF,~
$
\de_{\xi} A_0 =(A_0 \xi^0)_{,0}+\xi^k \nabla_k A_0+A_k \xi^k_{,0},
~\de_{\xi} A_i =A_{i,0}\xi^0+\xi^k \nabla_k A_i+A_k \nabla_i\xi^k,
$
as well as (\ref{delg3}) for the \Ho~gravity action (\ref{horava}),
where $E_i=\dot{A}_i -\nabla_i A_0, F_{ij}=\nabla_i A_j-\nabla_j A_i$,
and $\eta, \zeta$ are the electromagnetic coupling constants
\cite{Suda:1992,Kiri:2009,Lin:2014}.
Then, solving the full equations from variations of $A_0$ and $A_i$, as well as $N, N^i$, and $g^{ij}$ as in Sec. III (see Appendix {\bf E} for the details), we obtain the gauge potential, with electric charge $q_e$ and magnetic charge $q_m$,
\beq
A=-\sqrt{\f{\xi}{\eta}} \f{\left[ q_e r \De_\theta + q_m a ~\mbox{cos}\theta~ (1-\La r^2 (3 \xi)^{-1})\right]}{\rho^2 \Xi} dt+ \f{1}{\sqrt{\eta \ka}} \f{\left[q_e r {\it a}~ \mbox{sin}^2\theta+q_m (r^2+a^2)~ \mbox{cos}\theta \right]}{\rho^2 \Xi} d \phi
\label{A_sol}
\no \\
\eeq
and the metric (\ref{KAdS_ansatz}), (\ref{KAdS_sol}) by replacing
$2mr\ra2mr-(q_e^2+q_m^2)$, but {\it only with} the ``noble"
coupling,
\beq
\zeta \eta^{-1}=\ka \xi.
\label{noble_coupling}
\eeq
It is rather surprising that the exact charged black hole solutions exist only in the
noble coupling (\ref{noble_coupling}), due to coupling to \Ho~ gravity
\footnote{It is interesting that the noble coupling can be also obtained
{by}
a Kaluza-Klein reduction from $(4+1)$-dimensional {\it kinetic-conformal} pure \Ho~ gravity $(\la=1/4)$ \cite{Rest:2019}.}. Actually, this is the case where the speed of gravity is identical to the speed of light (or electromagnetism), $c_g=c_l=\sqrt{\ka \xi}$ with the dispersion relation $\om^2= (\ka \xi) \bf{k}^2$ for the perturbations around the flat Minkowski background \cite{Park:2009}.

The first law of thermodynamics (\ref{1st law}) is extended as (cf. \cite{Cald:1999}), with the same $S_H$ and $\Om_H$,
\beq
dM=T_H dS_H +\Om_{H} dJ+\Phi^{(e)}_H dQ_e+\Phi^{(m)}_H dQ_m
\label{1st law_charged}
\eeq
with the Hawking temperature $T_H$, the electric and magnetic chemical potentials $\Phi^{(e,m)}_H$,
\beq
T_H&=&\f{r_H \left(1-\f{a \La}{3 \xi}-\f{r_H^2 \La}{\xi}-\f{a^2+q_e^2+q_m^2}{r_H^2} \right)}{4 \pi(r_H^2+a^2)}, \no \\
\Phi^{(e)}_H&=&16 \pi \sqrt{\xi \eta}~\f{q_e r_H}{r_H^2+a^2}, ~
\Phi^{(m)}_H=16 \pi \xi \sqrt{\ka \eta}~\f{q_m r_H}{r_H^2+a^2},
\eeq
and the canonical charges, computed from fluxes of electromagnetic fields,
\beq
Q_e=\sqrt{\f{\xi}{\eta}}\f{q_e}{\Xi}, ~Q_m=\f{1}{\sqrt{\ka \eta}}\f{q_m}{\Xi},
\eeq
which are also {\it r-independent} \footnote{For {\it dyonic} black holes, electric (magnetic) canonical charge has ``$r-$dependent" and ``induced" magnetic (electric) charge contribution also due to rotation, but they are exactly canceled and have no effects in (\ref{1st law_charged}).}, as in $M$ and $J$ (\ref{M,J_KdS}) in gravity.
On the other hand, another standard form of rotating (A)dS black hole solutions
with a non-vanishing angular velocity $\Om_{\infty}$ at infinity
is obtained by
$\widehat{A}=A-({a \La}/{3 \xi}) A_{\phi}dt$,
considering a coordinate transformation $d \phi=d \widehat{\phi}-(a \La/3 \xi)dt$ also.

The ergo-sphere is larger $(\ka \xi >1)$ or smaller $(\ka \xi <1)$ than that of GR. In particular, for the \asy~flat case, the ergo-surface ($g_{tt}=0$) is given by
\beq
\mbox{sin}^2\theta=\f{
\left\{ (r^2+a^2)^2+\De_r^2+(\ka \xi-1) (2mr)^2
-\sqrt{\left[ \left(r^2+a^2+\De_r \right)^2+ (\ka \xi-1) (2mr)^2 \right] \ka \xi (2mr)^2} \right\}
}{2 a^2 \De_r},\no \\
\eeq
which exists {\it always} for $\ka \xi>0$, from $g_{tt}|_{r_+}=(\ka \xi) a^2 \mbox{sin}^2 \theta /\rho^{2}|_{r_+}>0$. If we assume the {\it area non-decreasing} case with the appropriate energy conditions \cite{Hawk:1971}, one can consider the Penrose process \cite{Penr:1969} to extract positive energy from inside of the ergo-surface ($g_{tt}< 0$) and obtain {\it formally} the same energy-radiating efficiency as in Kerr metric, due to the same area formula ${{\cal A}}=8 \pi m (m+\sqrt{m-a^2})$. However, its wave analogue, known as the {\it super-radiant} scattering is not quite clear, due to the lack of separability of wave equations (see Appendix {\bf D}).

\section{Observational constraints}

The recent observation of the arrival delay of $(+1.74 \pm 0.05)s$ in the  {\it coincident} gravitational waves (GW) and gamma rays (GW170817, GRB170817A) \cite{LIGO:2017} yields the constraints on the difference of the gravity speed $c_{g}$ and the gamma-ray photon's light speed $c_l$,
\beq
-3 \times 10^{-15} c_l < (c_g-c_l) < 7 \times 10^{-16} c_l,
\label{GRB}
\eeq
where the upper bound, or the lower bound, is obtained by assuming the simultaneous emission of the GW and GRB signals, or the GRB signals $10 s$ after the GW signal, respectively. But, not to mention the important change of the order of magnitudes by changing the models of photon emissions \cite{LIGO:2017}, we note that one can not tell about $c_g$ separately from the data (\ref{GRB}), without knowing about $c_l$ together, since (\ref{GRB}) is only about theirs difference $c_g-c_l$. The smallness of the relative difference $(c_g-c_l)/c_l$ in (\ref{GRB}) can be due to the smallness of $(\De c_g-\De c_l)/c_l$ for the modified speeds of gravity and light, $c_g=c+\De c_g, ~c_l=c+\De c_l,$ from the standard light speed $c$. And also, the observed time delay of the light can be due to the space-time curvature along its path.

In the literatures \cite{Emir:2017,Gong:2018}, on the other hand, it has always been assumed that $\De c_l=0$ and so (\ref{GRB}) has been considered as the absolute bound of $-3 \times 10^{-15}< \De c_g/c < 7 \times 10^{-16}$, which would be strong enough to exclude the modified speed of gravity $\De c_g$. However, if there is $\De c_l$ also, the story is completely different and the constraint (\ref{GRB}) yields $-3 \times 10^{-15}< (\De c_g-\De c_l)/c < 7 \times 10^{-16}$, even if $\De c_g$ and $\De c_l$ are not small separately! Actually, our exact solution for charged rotating black holes implies that $c_g=c_l$, {\it i.e.}, $\De c_g=\De c_l$ for the small perturbations around the flat Minkowski background \cite{Park:2009}. In that case, the time delay will be due to either (a) the emission-time difference of the GW and GRB signals, or (b) due to the photon's mass, either intrinsic or curvature induced, or even (c) the graviton's mass which has not been studied in our solutions. Hence, our LV gravity with $c_g \neq c$ can be still consistent with the recently observed time delay of the coincident GW and GRB signals from a binary neutron star merger, contrary to the claims in the literature, with some possible scenarios (a)-(c) that we listed above, to explain the observed data.

Another important constraints are the parameterized post-Newtonian (PPN) parameters which measure the preferred-frame effects \cite{Blas:2011,Emir:2017}, with the current bounds \cite{Will:2014},
\beq
|\al_1|&=&|8 (\widehat{\xi}-\si/2)| < 4 \times 10^{-5},\label{alpha_1} \\
|\al_2|&=&\left|\left(\f{\widehat{\xi}-\si/2}{\widehat{\xi}-\si/2+1}\right)
\left(1+\f{2 (2 \widehat{\la}+1) (\widehat{\xi}-\si/2)}{\widehat{\la}} \right)\right| < \times 10^{-9},\label{alpha_2}
\eeq
where $\widehat{\xi}=\xi-1$ and $\widehat{\la}=\la-1$.
For the case of $\widehat{\xi}-\si/2=0$, which includes the GR limit of
$\widehat{\xi}=0, \si=0$, we obtain $\al_1=\al_2=0$ which trivially satisfies
the constraints (\ref{alpha_1}) and (\ref{alpha_2}), without any restriction on
$\widehat{\la}$. However, for $\widehat{\xi}-\si/2\neq 0$, the situation is quite
different since, in the constraint (\ref{alpha_2}), $\widehat{\la}$ can not
contain the GR case of $\widehat{\la}=\la-1=0$ because it makes $\al_2\ra \infty$.
Rather, it
has a gap
$\widehat{\la} \approx -2 (\widehat{\xi}-\si/2)+\widehat{\ep}$ with $\widehat{\ep}>2 \times 10^{-4}$ using $2 \widehat{\la} +1 \approx 1$ for a small $\widehat{\la}$, with the constraints $|\widehat{\xi}-\si/2|< 5 \times 10^{-6}$ from (\ref{alpha_1}). In other words, the LV parameters $\widehat{\xi}-\si/2$ and $\widehat{\la}$ are strongly correlated in \Ho~ gravity.

\section{Concluding remarks}

In conclusion, we have studied a procedure for finding rotating black hole solutions in the low-energy sector of the four-dimensional (non-projectable) \Ho~gravity as a viable LV gravity, and obtain the Kerr-type rotating black hole solutions with or without cosmological and electromagnetic charges.

It would be interesting to test the solution in the astrophysical observations. Applying our procedure to the renormalizable $(z=3)$ \Ho~gravity in four dimensions and finding the exact rotating black hole solutions will a challenging problem. The uniqueness of our solutions and their stabilities are remaining problems. Especially, it would be interesting to see the instability of {\it ultra-spinning} black holes $J/aM=1/\sqrt{\ka \xi}>1$ with $\ka \xi < 1$, similar to ultra-spinning Myers-Perry's black holes in GR \cite{Empa:2003}.\\

{\it Note added}: After finishing this paper, a related paper, which corresponds to $\ka \xi=1$, $\la \neq 1$ in our context, appeared in Einstein-aether gravity \cite{Fran:2023}, where the numerical rotating solutions have been found recently \cite{Adam:2021}. But the possibilities of our rotating solutions with $\ka \xi \neq 1$ were not considered due to the recent gravity speed constraints \cite{LIGO:2017} without considering the possibility of modified light speed \cite{Emir:2017,Gong:2018}.

\section*{Acknowledgments}

We would like to thank {\"O}zg{\"u}r Sar{\i}o\u{g}lu
for helpful correspondences.
This work was supported by Basic Science Research Program through the National
Research Foundation of Korea (NRF) funded by the Ministry of Education,
Science and Technology {(2020R1A2C1010372, 2020R1A6A1A03047877)}.

\appendix

\section{Computational details on mass and angular momentum}

In this Appendix, we present computational details of the
mass and angular momentum (\ref{M,J}) or (\ref{M,J_KdS}) found from both the standard
canonical approach \cite{Bana:1992} and the recently-proposed covariant formalism
\cite{Deve:2021}.

\subsection{Standard canonical approach}

In the standard canonical approach \cite{Bana:1992}, we first need to {\it recover} the neglected integration constants in our ansatz
(\ref{massive_Kerr_metric}), (\ref{massive_Kerr_functions})
by considering the generalized ansatz
\beq
ds^2=-\widetilde{N}^2 c^2 dt^2+g_{ij}\left(dx^i+\widetilde{N}^i dt\right)\left(dx^j+\widetilde{N}^jdt\right),
\label{metric_general}
\eeq
with $\widetilde{N}=W(\infty) N,~\widetilde{N}^i=W(\infty) N^i+N^i (\infty)$ so that the ansatz (\ref{massive_Kerr_metric}), (\ref{massive_Kerr_functions}) corresponds to the usual choice of $W(\infty)=1, N^i (\infty)=0$: One can easily check that the generalized ansatz with the solution (\ref{f,g_sol}) is also the solution of (\ref{eom1})-(\ref{eom3}), which is due to the unchanging action with the generalized metric (\ref{metric_general}).

By plugging the generalized ansatz (\ref{metric_general}) into the first-order action form of (\ref{horava})
\beq
S_g=\int dt dx^3 \left(\widetilde{\pi}^{ij} \dot{g}_{ij}-\widetilde{N}\widetilde{\cal H} -\widetilde{N}^i\widetilde{\cal H}_i \right)
\label{S_g_1st_order}
\eeq
such that $S=S_g+{\cal B}$ is {\it differentiable} with the appropriate boundary action ${\cal B}$. Here, $\widetilde{\pi}^{ij}=(1/\ka) \sqrt{g} (\widetilde{K}^{ij}-\la \widetilde{K} g^{ij})$, $\widetilde{K}^{ij}=(2 \widetilde{N})^{-1} (\dot{g}_{ij}-\nabla_i \widetilde{N}_j-\nabla_j \widetilde{N}_i)$, $\widetilde{\cal H}$, and $\widetilde{\cal H}_i$ are the canonical momenta, extrinsic curvature, Hamiltonian constraint, and momentum constraints for the generalized metric (\ref{metric_general}), respectively.

Then, for the stationary ansatz, {\it i.e.}, $\dot{g}_{ij}=0$, the canonical mass $M$ and angular momentum $J$ are defined as the conjugates to $W(\infty)$ and $N^{\phi}(\infty)$, respectively, in the variation of $\delta S_g$,
\beq
\delta S_g =(bulk~ terms)+(t_2-t_1) \left[W(\infty) \delta M - N^{\phi}(\infty) \delta J \right]
\label{A3}
\eeq
such that the {\it non-differentiable} boundary variations in (\ref{A3}) are canceled
by the variation ${\de\cal B}$ of
\beq
{\cal B}=(t_2-t_1) [-W(\infty) M+N^{\phi} (\infty) J ]
\eeq
with the boundary conditions $\de N^{\phi} (\infty)=\de W(\infty)=0$.

By plugging our ansatz (\ref{massive_Kerr_metric}), (\ref{massive_Kerr_functions}) and considering the $r\ra \infty$ limit, after the time and angular integrations, we finally obtain the boundary variation terms
\beq
\delta S_g =(bulk~ terms)+(t_2-t_1) 16 \pi \left[\f{W(\infty)}{\Xi^2}\xi \delta m- \f{N^{\phi}(\infty)}{\Xi^2} \sqrt{\f{\xi}{\ka}}\delta (a m) \right]
\label{A5}
\eeq
such that, be comparing (\ref{A5}) with (\ref{A3}), we obtain
\beq
M=\f{16 \pi \xi m}{\Xi^2}, ~J=\f{16 \pi a m \sqrt{\xi \ka^{-1}}}{\Xi^2},
\label{M,J_KdS_standard}
\eeq
which gives (\ref{M,J_KdS}), or (\ref{M,J}) for asymptotically flat case.
Here, the first and second boundary terms in (\ref{A5}) come from
$-\widetilde{N}\de \widetilde{\cal H}$ and $-\widetilde{N}^i\de \widetilde{\cal H}_i$ of $\de S_g$ in (\ref{S_g_1st_order}), respectively, whose explicit forms are omitted due to their messy expressions. In this approach, it is important that we need to treat the parameters $m$ and $a$ as
variables in the variation (\ref{A5}), but treat them as the constants parameters
only after getting (\ref{M,J_KdS_standard}). For the boundary at $r\ra \infty$ in
de Sitter space ($\La>0$), (\ref{M,J_KdS_standard}) corresponds to the
conserved charges in $dS/CFT$ \cite{Park:1998}.

\subsection{Covariant formalism}

In the recently-proposed covariant formalism \cite{Deve:2021}, it was shown that the Ho\v{r}ava gravity action \footnote{In \cite{Deve:2021}, a more general action with (spatially) higher-curvature potential is used in the proof.} (\ref{horava}) is invariant under the (full) ${\it Diff}$ generated by a vector $\xi^{\mu}$ with a certain condition called ``supercondition". Noether theorem then guarantees the following conserved currents
\begin{\eq}
{\bf \cal{J}}^{\mu} (\de_{\xi} g)
&=& \Theta^{\mu} -\xi^{\mu} {\cal L}-{\bf \Sigma}^{\mu}-\partial_{\nu} {\cal U}^{\mu \nu},\label{Noether}
\end{\eq}
where $\Theta^{\mu}$ is the boundary term in $\de_\xi S_g$, ${\cal L}$ is the Lagrangian
density with the action
$S_g=\int dt dx^3 {\cal L}$,
$\Sigma^{\mu}$ is due to the apparent noncovariance of $S_g$,
and ${\cal U}^{\mu \nu}=-{\cal U}^{\mu \nu}$ is called
{\it super-potential} (see \cite{Deve:2021} for their explicit forms). From the
covariant conservation law $\nabla_\mu {J}^{\mu}=0$ with
${J}^{\mu} =(\sqrt{g} N)^{-1}{\bf \cal{J}}^{\mu}$ and Stokes theorem, we can
define the
conserved charge ($\Theta^0=\Sigma^0=0$)
\beq
Q&=&\int d^{3}x \sqrt{g}\,n_{\mu}{J}^{\mu},\\
&={}&- \int d^{3}x \sqrt{g}\,N\,{J}^{0},\\
&={}& \int d^{3}x (\xi^{0}{\cal L}+\partial_{i} {\cal U}^{0 i}), \label{Q}
\eeq
where
\beq
{\cal L}&=&\frac{1}{\kappa}(K_{ij}K^{ij}-\lambda K^{2})+\xi R -2 \Lambda ,\\
{\cal U}^{0 j}&={}&\frac{2}{\kappa}\left(\xi^{0}N_{i}+\xi_{i}\right)\left(K^{ij}-\lambda g^{ij}K\right)
\eeq
for the Killing vector $\xi^{\mu}$ \footnote{We can choose the zeroth component of the Killing vector as $\xi^{0}=-1$ for the energy (or mass) and space components of the vector $\xi^{i}=\delta^{i}_{\phi}$ for the angular momentum.}.

At this point, we first need to change (\ref{Q}) into the total derivatives plus the bulk terms, and then compute the charge by plugging
our ansatz (\ref{massive_Kerr_metric}), (\ref{massive_Kerr_functions}),
as discussed in \cite{Deve:2021}. One simple way to achieve this is to leave $f(r)$ as the only unknown function in the ansatz (\ref{massive_Kerr_metric}), (\ref{massive_Kerr_functions}), and integrate by parts to get
\beq
Q=\int dr \, d\theta \, d\phi \left[\partial_{r}A(r,\theta)+B(r,\theta) \right],
\label{Q_Noether}
\eeq
where
\beq
A(r,\theta)&=&\frac{\xi a^2  {\rm sin}^3\theta  \left(1-r^2 \La/3\xi \right) f'(r) \rho^2 (r,\theta )-2  \xi {\rm sin}\theta ~r f(r) \left[f(r)+\Delta(r)+\Lambda \rho^4 (r,\theta )/3\xi\right]}{\Xi ^2 \rho^4 (r,\theta )}\xi^{0}\nonumber\\
&+&\frac{\sqrt{\xi } a  {\rm sin}^3\theta \left\{2 r f(r) \left[2 \rho^2 (r,\theta )+a^2  {\rm sin}^2\theta\right]-\left(a^2+r^2\right) f'(r) \rho^2 (r,\theta )\right\}}{\sqrt{\kappa }  \Xi ^2 \rho^4 (r,\theta )}\xi^{\phi},\\
B(r,\theta)&=&\f{B_{1}(r,\theta)}{ \Xi ^2 \rho^6 (r,\theta )}\xi^{0}+\f{B_{2}(r,\theta)}{\sqrt{\kappa }  \Xi ^2 \rho^6 (r,\theta )}\xi^{\phi},\\
B_{1}(r,\theta)&=&-2 \xi a^2  {\rm sin}\theta  f(r) \Big\{\left(3 {\rm sin}^2\theta-2\right) r^2+a^4 (\Lambda/3 \xi)  \left({\rm sin}^2\theta-2\right) {\rm cos}^4\theta \no \\
&&-a^2 {\rm cos}^2\theta \left[ 2+{\rm sin}^2\theta-(\Lambda/3 \xi)  \left(5 {\rm sin}^2\theta-2\right) r^2\right]\Big\},\nonumber\\
B_{2}(r,\theta)&=&-2 \sqrt{\xi }{a}^3 {\rm sin}^3\theta  f(r) \left[{a}^2 {\rm cos}^2\theta \left(4- {\rm sin}^2\theta\right)+\left(4-5 {\rm sin}^2\theta\right) r^2\right]\nonumber.
\eeq

By plugging our solution for $f(r)$ in (\ref{f,g_sol}), performing the angular integrations, and taking the
$r \rightarrow\infty$ limit, the $r$-dependent pieces from the boundary and bulk terms {\it vanish separately}, and we finally have
\beq
Q&=&-\xi^0 \f{16 \pi \xi m}{\Xi^2}+\xi^\phi \f{16 \pi a m \sqrt{\xi \ka^{-1}}}{\Xi^2}\no \\
&=&-\xi^0 M +\xi^\phi J,
\label{M,J,Noether}
\eeq
which agrees with (\ref{M,J_KdS}) or (\ref{M,J}).

Here, it is important to note that we would get the vanishing charges if we {\it first} plug our solutions (\ref{f,g_sol}) into the Noether charge formula (\ref{Q}) or (\ref{Q_Noether}) {\it before} integrations, analogous to the null results in standard variational approach with the constants $m$ and $a$ in the previous subsection. Actually, the non-vanishing charge $Q$ from the first integration process indicates its integration nature: If there is a (integration) constant $C\equiv Q/4 \pi$ in the boundary piece $A$, {\it i.e.}, $A(r, \theta)=C +(non-constant~ terms)$, the constant can only be obtained from the first integration process.

Moreover, it is important to note also that the $r$-dependent pieces from the boundary and bulk terms {\it cancel exactly} for arbitrary boundary radius $r$, only if the Killing vectors exist inside the boundary so that our conserved charges (\ref{M,J,Noether}) do not depend on the boundary hypersurface, regardless of cosmological constant $\La$! (See \cite{Deru:2004} for similar results in covariant theories) This remarkable property is not clear in the standard variational approach (\ref{A3}) which defines the conserved charges $M$ and $J$ only from the boundary variation terms, without its bulk companion, {\it i.e.} $B$ for $\pa_r A$ in (\ref{Q_Noether}).

\section{Explicit expressions of $\widehat{\rho}$ and $\widehat{p_a}$ in
${T^{\m}}_{\n (\it{eff})}$
}

The explicit expressions of $\widehat{\rho}$ and $\widehat{p_a}$ in
the effective energy-momentum tensor ${T^{\m}}_{\n (\it{eff})}$ (\ref{T_eff}) are given as follows:
\beq
\widehat{\rho} &=& \widehat{\kappa\xi}~
a^2 m^2  \text{sin}^2\theta
 \left\{
(a^4-2 a^2 r^2-3 r^4)^2-2 a^2 (a^6-5 a^4 r^2+a^2 (4 m-3 r)
r^3+3 r^6) \text{sin}^2\theta
\right.  \no \\
&&\left.
+a^4 (a^4-6 a^2 r^2+(8 m-3 r) r^3) \text{sin}^4\theta
 \right\}/Z, \\
\widehat{p_1} &=& \widehat{\kappa\xi} ~
a^2 m^2  \text{sin}^2\theta \left\{(a^4-2 a^2 r^2-3 r^4)^2-2 a^2 (a^6-a^4 r^2+3 r^6+a^2 r^3 (-4
m+r)) \text{sin}^2\theta
\right.  \no \\
&&\left.
+a^4 (a^4+2 a^2 r^2+(-8 m+5 r)r^3 ) \text{sin}^4\theta\right\}/Z ,\\
\widehat{p_2} &=&2 \widehat{\kappa\xi} ~
a^4 m^2 r \text{cos}\theta (a^2+r (-2 m+r)) \left[a^4-3 a^2 r^2-6 r^4+a^2 (a^2-r^2) \text{cos}(2 \theta)\right] \text{sin}^3\theta/Z, \\
\widehat{p_3} &=&2 \widehat{\kappa\xi}~a^4 m^2 r
\text{cos}\theta \left[a^4-3 a^2 r^2-6 r^4+a^2 (a^2-r^2) \text{cos}(2 \theta )\right]
\text{sin}^3\theta/Z, \\
\widehat{p_4} &=&8 \sqrt{\kappa \xi} ~\widehat{\rho}~ {\it a} m r/\Si^2 ,\\
Z&=&(a^2+r^2-a^2 \text{sin}^2\theta)^3 \left\{(a^2+r^2)^2-a^2 (a^2+r (-2 m+r)) \text{sin}^2\theta\right\}^2,
\eeq
where $\widehat{\kappa\xi} =\kappa  \xi-1$ is a LV
factor.

\section{The uniqueness of the invariant line element $ds^2$ under ${\it Diff}_{\cal F}$.}

In
order to prove the {\it uniqueness} of the invariant line element (\ref{metric})
under
${\it Diff}_{\cal F}$, we start by expanding (\ref{metric}) as
\beq
ds^{2}=(-N^{2}+g_{ij}N^{i}N^{j})dt^{2}+2 g_{ij}N^{i}dx^{j}dt+g_{ij}dx^{i}dx^{j}.
\eeq
Then, under ${\it Diff}_{\cal F}$ (\ref{delg3}), the first term transforms as
\beq
\delta_{\xi}\left[ (-N^{2}+g_{ij}N^{i}N^{j})dt^{2}\right]&=&(\xi^{0}\partial_{0}g_{tt}+\xi^{i}
\nabla_i g_{tt}+\underline{2N_{i}\partial_{0}\xi^{i}})dt^{2},\label{D1}
\eeq
and the second and third terms transform, respectively, as
\beq
\delta_{\xi} \left[2 g_{ij}N^{i}dx^{j}dt\right]&=& -\underline{2N_{j}\partial_0\xi^{j}}dt^{2}+(2\xi^{0}\partial_0N_{j}+2\xi^{i}\nabla_{i}N_{j}
+2N^{i}\nabla_{j}\xi_{i} \no \\
&&-2N_{k}\partial_{j}\xi^{k}+\underline{\underline{2g_{ij}\partial_0\xi^{i}}})dtdx^{j},\label{D2}\\
\delta_{\xi} \left[g_{ij}dx^{i}dx^{j}\right]&=&(2\nabla_{i}\xi_{j}+\xi^{0}\partial_0g_{ij}
-2g_{ik}\partial_{j}\xi^{k})dx^{i}dx^{j}-\underline{\underline{2g_{ij}\partial_0\xi^{j}}}dx^{i}dt.\label{D3}
\eeq
Summing up (\ref{D1}-\ref{D3}) and using the metric compatibility to change the covariant derivatives to the partial ones,
we have the form of Lie derivative of a scalar as follows:
\beq
\delta_{\xi}(ds^{2})&=&(\xi^{0}\partial_0g_{tt}+\xi^{i}\partial_{i}g_{tt})dt^{2}
+(2\xi^{0}\partial_0N_{j}+2\xi^{i}\partial_{i}N_{j})dt dx^{j} \nonumber\\
&&+(\xi^{0}\partial_0g_{ij}+\xi^{k}\partial_{k}g_{ij})dx^{i}dx^{j}\\
&=&{}\xi^{\mu} \partial_{\mu} (ds^{2}).
\eeq
Here, it is important to note that the {\it underlined} terms are
{\it exactly canceled} only for the line element (\ref{metric}), which means
the unique line element that is invariant under ${\it Diff}_{\cal F}$ (\ref{delg3}).
This is quite remarkable since it seems to be contradict to the transformation
of action (\ref{horava}), in which each term is invariant {\it separately}
under ${\it Diff}_{\cal F}$ so that one can introduce arbitrary coupling
constant for each term as the LV
effect. The basic reason is that $dx^i$ is $not$ a spacial projection of $dx^{\mu}$, but rather it can transform to $dt$ also (though not for the reverse) when $\partial_0 \xi^i \neq 0$, contrary to the spatially-projected quantities $K_{ij}, R_{ij}, \nabla_i, etc$ in the action (\ref{horava}). This is a quite powerful property of our LV gravity that enables us to study invariant properties of particles via $ds^{2}$ also as in GR, contrary to widespread beliefs. Our proof, which has never been known in literature, justifies the use of $ds^{2}$ even in \Ho~gravity case also, where the full {\it Diff} is apparently broken.

\section{Petrov classification, Killing tensor, and Hamilton-Jacobi equation}

For the Petrov classification of our rotating solution (\ref{massive_Kerr_metric}) without cosmological constant, we first find the four {\it null} vectors,
\beq
\widehat{l}^{\m}&=&\f{1}{\De_r} \left[r^2+a^2,\De_r,0,a+ (r^2+a^2) {\it H}\right], \no \\
\widehat{n}^{\m}&=&\f{1}{2 \rho^2} \left[r^2+a^2,-\De_r,0,a+ (r^2+a^2) {\it H}\right], \no \\
\widehat{m}^{\m} &=&\f{1}{\sqrt{2} (r+i {\it a} \mbox{cos}\theta)} \left[i {\it a } \mbox{sin}\theta,0,1,i+i{\it a} \mbox{sin}\theta {\it H} \right],
\label{null_new}
\eeq
which satisfy the usual orthogonality conditions $\widehat{l}^{\m} \widehat{n}_{\m}=-1, \widehat{m}^{\m} \widehat{\bar{m}}_{\m}=1$ with ${\it H}=(\sqrt{\ka \xi}-1) 2mra/\Si^2$, and our metric can be written as $g_{\m \n}=2 \widehat{m}_{(\m} \widehat{\bar{m}}_{\n)}-2 \widehat{l}_{(\m} \widehat{n}_{\n)}$, as usual. But, we find that there are {\it four} distinct principal null directions $k^{\m}$
that satisfy $k^{\al}k^{\be}k_{[\m} C_{\n ] \al \be [\rho} k_{\si]}=0$ with the Weyl tensor $C_{\n \al \be \rho}$, such that our solution is the Petrov type I \footnote{Similar results for the perturbed spacetimes are also obtained recently \cite{Aran:2015,Owen:2021}.}. This is contrary to Kerr spacetime in GR, where two null vectors $l^{\m}$ and $n^{\n}$ are aligned with the principal null direction and it is the Petrov type D.

Since our rotating solution (\ref{massive_Kerr_metric}) is the Petrov type I,
several well-established and nice properties of type D are not guaranteed. For example, the existence of Killing tensor and also the separability of Hamilton-Jacobi (HJ) equation are not guaranteed \cite{Cart:1973,Walk:1970}. Especially for Carter's proof, our solution can {\it not} be expressed as the generic form of solutions which generates the Killing tensor and separable HJ equation \cite{Cart:1973}.

Actually one can check that the {\it used-to-be} Killing tensor form of Kerr metric $L_{\m \n}=2 \rho^2 m_{(\m} \bar{m}_{\n)}-a^2 \mbox{cos}^2 \theta g^{(4)}_{\m \n}$ with the complex null vectors $m_{\m}, \bar{m}_{\n}$ of the Newman-Penrose basis in GR, nor its correspondent in our new null vectors (\ref{null_new}) $\widehat{L}_{\m \n}=2 \rho^2 \widehat{m}_{(\m} \widehat{\bar{m}}_{\n)}-a^2 \mbox{cos}^2 \theta g^{(4)}_{\m \n}$, do not satisfy the Killing tensor equation $\nabla^{(4)}_{(\m} L^{}_{\n \si)}=0$ or $\nabla^{(4)}_{(\m} \widehat{L}^{}_{\n \si)}=0$. However, interestingly, we can find a certain geodesic with tangent vector $v^{\m}$ such that $v^{\n} v^{\si}L_{\n \si}$ is still {\it constant} along the geodesics, {\it i.e.},
\beq
v^{\m} \nabla^{(4)}_{\m} (v^{\n}v^{\si} L_{\n \si})=-\f{24 (\ka \xi-1) (mra^2)^2~ v_{\theta} v_{\phi}^2~ \mbox{sin}\theta \mbox{cos}\theta}{\Si^4}=0.
\label{Killing_condition}
\eeq
The fixed-angle geodesics, $v_{\theta}=0$ ($\theta$=const.) or $v_{\phi}=0$ ($\phi$=const.) are those cases and this supports our previous analysis of $\theta=\pi/2$ plane in Sec. VI. For other more general geodesics with $v_{\theta}v_{\phi} \neq 0$, we may need numerical analysis to integrate the geodesics completely if there is no exact Killing tensor \footnote{If one considers successive applications of infinitesimal geodesics with fixed-angle planes, we may produce  general geodesics in the continuous limit. The condition (\ref{Killing_condition}), then, can be generalized into an integral form $\int_{\ga}v^{\m} \nabla^{(4)}_{\m} (v^{\n}v^{\si} L_{\n \si})=0$ along a general geodesic $\ga$.}.

A related difficulty in integrating the full geodesics is the {\it lack of separability} of HJ equation,
\beq
-2 \f{\pa S}{\pa \tau}&=&g^{(4)\m \n} \pa_{\m}S \pa_{\n}S \no \\
&=&-\f{\Si^2}{\rho^2 \De_r}(\pa_t S)^2+\f{4  mra \sqrt{\ka \xi}}{\rho^2 \De_r}\pa_t S \pa_{\phi}S+\f{\De_r \rho^4 - (2mra)^2  \ka \xi \mbox{sin}^2\theta}{\Si^2 \rho^2 \De_r \mbox{sin}^2\theta}(\pa_{\phi}S)^2 \no \\
&&+\f{\De_r}{\rho^2}(\pa_r S)^2+\f{1}{\rho^2}(\pa_{\theta} S)^2
\label{HJ}
\eeq
from a particle Hamiltonian $H=(1/2)g^{\m\n}p_{\m} p_{\n}$ and $p_{\m}=\pa_{\m}S$ with the Hamilton's principal function $S$. By ``assuming" the separability, we consider \cite{Chan:1985}
\beq
S=\f{1}{2} \m^2 \tau -Et+L \phi+S_r (r)+S_{\theta} (\theta)
\eeq
then, we finally obtain the master equation
\beq
\left\{
\m^2 r^2 +\left(a E+\sqrt{\ka \xi} L \right)^2-\f{1}{\De_r} \left((r^2+a^2)E +\sqrt{\ka \xi} a L \right)^2+\De_r \left( \f{d S_r}{dr}\right)^2
+(\ka\xi-1) \f{a^2 L^2}{\De_r}
\right\} \no \\
+\left\{
\m^2 a^2 \mbox{cos}^2\theta+ \left(-a^2 E^2+\f{\ka \xi L^2}{\mbox{sin}^2\theta} \right)
+\left( \f{d S_{\theta}}{d\theta} \right)^2 -(\ka\xi-1) \f{L^2}{\mbox{sin}^2\theta}
\right\}
+(\ka\xi-1) \f{(2mra)^2 L^2}{\Si^2 \De_r}=0,\no \\
\label{HJ_master}
\eeq
where $\m$ is the particle's rest mass. The last term in (\ref{HJ_master}) breaks the separability of HJ equation in our LV gravity with $\ka \xi \neq 1$, unless we consider $L=0$ or $\theta$=constant, similar to the above Killing tensor analysis.

\section{Equations for the Lorentz-violating Maxwell action}

In order to find the LV Maxwell action, we first consider
the covariant Maxwell action,
\beq
S_{M_0}=\eta\int d^{4}x \sqrt{-g^{(4)}}  F_{\mu\nu}F^{\mu\nu}
\label{covmax}
\eeq
with the electromagnetic coupling constant $\eta$.
Applying the usual ADM decomposition to this action and, due to ${\it Diff}_{\cal F}$ (\ref{delg3}), introducing
an arbitrary coupling
 parameter $\zeta$ where the covariant case (\ref{covmax}) corresponds to $\zeta=\eta$, we obtain the LV Maxwell action,
\beq
S_{M}=\int_{{\bf R}\times \Si_t} dt d^{3}x \sqrt{g} N \left(\zeta F_{ij}F^{ij}\mp \frac{2\eta}{N^{2}}\left[E_{i}+F_{ij}N^{j}\right]^{2}\right),
\label{LV_Max-action}
\eeq
where $\mp$ corresponds to a normalization of the normal vector
$n_{\mu}n^{\mu}=\mp 1$ for a
hypersurface $\Si_t$,
and $E_{i}\equiv
\dot{A}_{i}-\partial_{i}A_{0},~ F_{ij}=\nabla_i A_j-\nabla_j A_i$.

The variations of the LV Maxwell action (\ref{LV_Max-action}) give, neglecting boundary terms and choosing the hypersurface $\Si_t$ as a time-foliation with $n_{\mu}n^{\mu}=-1$,
\beq
{\cal H}_M\equiv {-}\frac{\delta S_{M}}{\delta N}&=&{-}\zeta F_{ij}F^{ij} {-}2 \eta L_{i}L^{i},\\
{\cal H}^i_{M}\equiv  {-}\frac{\delta S_{M}}{\delta N_{i}}&=& {-}4\eta L_{j}F^{ji}, \\
E_{ij}^M \equiv \frac{\delta S_{M}}{\delta g^{ij}}&=&-\frac{1}{2}N\mathcal{L}_M g_{ij}+2\zeta N F_{i}{}^{k}F_{kj}-4\eta  L^{k}F_{k(j}N_{i)}-2 \eta N L_{i}L_{j},\\
E_{0}^M \equiv \frac{\delta S_{M}}{\delta A_{0}}&=&\eta \nabla^{i}L_{i}=0,\\
E_{i}^M \equiv\frac{\delta S_{M}}{\delta A_{i}}&=& -\zeta \nabla_{j}(4N F^{ji})+4 \eta \dot L^{i}+4 \eta \nabla_{j}\left(L^{j}N^{i}-L^{i}N^{j}\right)=0,\\
\eeq
where $\mathcal{L}_M=\zeta F_{ij}F^{ij}-2 \eta L_{i}L^{i} $ and $L_{i}\equiv N^{-1} \left(E_{i}+F_{ij}N^{j}\right)$. Then, the total gravity equations are given by ${\cal H}_{tot}={\cal H}+{\cal H}_M=0,~ {\cal H}^i_{tot}={\cal H}^i+{\cal H}^i_{M}=0,$ and ${E_{ij}}^{tot}=E_{ij}+E_{ij}^M=0.$

\newcommand{\J}[4]{#1 {\bf #2} #3 (#4)}
\newcommand{\andJ}[3]{{\bf #1} (#2) #3}
\newcommand{\AP}{Ann. Phys. (N.Y.)}
\newcommand{\MPL}{Mod. Phys. Lett.}
\newcommand{\NP}{Nucl. Phys.}
\newcommand{\PL}{Phys. Lett.}
\newcommand{\PR}{Phys. Rev. D}
\newcommand{\PRL}{Phys. Rev. Lett.}
\newcommand{\PTP}{Prog. Theor. Phys.}
\newcommand{\hep}[1]{ hep-th/{#1}}
\newcommand{\hepp}[1]{ hep-ph/{#1}}
\newcommand{\hepg}[1]{ gr-qc/{#1}}
\newcommand{\bi}{ \bibitem}


\begin{thebibliography}{999}


\bibitem{Kerr:1963}
  R.~P.~Kerr,
  Phys.\ Rev.\ Lett.\  {\bf 11}, 237 (1963).

\bibitem{Chan:1985}
S.~Chandrasekhar,
``The mathematical theory of black holes,'' (Oxford Univ.
Press, 1983).


\bibitem{Cart:1973}
B.~Carter, 
in {\it Les Astre Occlus}, Gordon and Breach, New York (1973).


\bibitem{Walk:1970}
M.~Walker and R.~Penrose,
Commun. Math. Phys. \textbf{18}, 265
(1970).

\bibitem{Kerr:1965}
R.~P.~Kerr and A.~Schild,
Gen. Rel. Grav. \textbf{41},
2485
(2009).

\bibitem{DeWi:1967}
  B.~S.~DeWitt,
  Phys.\ Rev.\  {\bf 160}, 1113 (1967).

\bibitem{Hora:2009} P.~Ho\v{r}ava,
  Phys.\ Rev.\  D {\bf 79}, 084008 (2009).

\bibitem{Park:2009}
M.~I.~Park,
Class. Quant. Grav. \textbf{28}, 015004 (2011).


\bibitem{Blas:2009}
D.~Blas, O.~Pujolas and S.~Sibiryakov,
Phys. Rev. Lett. \textbf{104}, 181302 (2010).

\bibitem{Blas:2009b}
D.~Blas, O.~Pujolas and S.~Sibiryakov,
JHEP \textbf{10}, 029 (2009).

\bibitem{Bell:2010}
J.~Bellorin and A.~Restuccia,
Int. J. Mod. Phys. D \textbf{21}, 1250029 (2012).

\bibitem{Deve:2020}
D.~O.~Devecioglu and M.~I.~Park,
Eur. Phys. J. C \textbf{80},
597 (2020)
[Erratum: Eur. Phys. J. C \textbf{80},
764 (2020)].


\bibitem{Jaco:2013}
T.~Jacobson,
Phys. Rev. D \textbf{89}, 081501 (2014).

\bibitem{Blas:2011}
D.~Blas and H.~Sanctuary,
Phys. Rev. D \textbf{84}, 064004 (2011).


\bibitem{Park:2023}
M.~I.~Park and H.~W.~Lee,
arXiv:2309.13859 [hep-th].


\bibitem{Park:2001}
M.~I.~Park,
Nucl. Phys. B \textbf{634}, 339
(2002).

\bibitem{Gibb:1977}
G.~W.~Gibbons and S.~W.~Hawking,
Phys. Rev. D \textbf{15}, 2738
(1977).

\bibitem{Gibb:2017}
G.~W.~Gibbons and M.~S.~Volkov,
Phys. Rev. D \textbf{96},
024053 (2017).

\bibitem{Bana:1992}
T.~Regge and C.~Teitelboim,
Annals Phys. \textbf{88}, 286 (1974);
M.~Banados \textit{et al.},
Phys. Rev. D \textbf{48}, 1506
(1993).

\bibitem{Deve:2021}
D.~O.~Devecioglu and M.~I.~Park,
Phys. Rev. D \textbf{108}, 
064030 (2023).

\bibitem{Lu:2009}
  H.~Lu, J.~Mei and C.~N.~Pope,
Phys.\ Rev.\ Lett.\  {\bf 103}, 091301 (2009);
  A.~Kehagias and K.~Sfetsos,
Phys.\ Lett.\ B {\bf 678}, 123 (2009);
  M.~I.~Park,
  JHEP {\bf 0909}, 123 (2009);
E.~B.~Kiritsis and G.~Kofinas,
JHEP \textbf{01}, 122 (2010).

\bibitem{Park:2012}
M.~I.~Park,
Phys. Lett. B \textbf{718}, 1137
(2013) [Erratum: Phys. Lett. B \textbf{809}, 135720 (2020)];
PTEP \textbf{2022}, 
113E02 (2022);
T.~P.~Sotiriou, I.~Vega and D.~Vernieri,
Phys. Rev. D \textbf{90}, 
044046 (2014).

\bibitem{Gibb:2004}
G.~W.~Gibbons \textit{et al.},
J. Geom. Phys. \textbf{53}, 49
(2005), Appendix {\bf E}.

\bibitem{Curi:1979}
A.~Curir,
Il Nuovo Cimento {\bf B52}, 262
(1979).

\bibitem{Klemm:2004}
D.~Klemm and L.~Vanzo,
JCAP \textbf{11}, 006 (2004).

\bibitem{Park:2006}
M.~I.~Park,
Phys. Lett. B \textbf{647}, 472
(2007);
Phys. Lett. B \textbf{663}, 259
(2008);
Class. Quant. Grav. \textbf{25}, 095013 (2008).

\bibitem{Cveti:2018}
M.~Cveti\v{c} \textit{et al.},
Phys. Rev. D \textbf{98},
106015 (2018);
T.~Jacobson and M.~Visser,
SciPost Phys. \textbf{7},
079 (2019).

\bibitem{Hawk:1998}
S.~W.~Hawking, C.~J.~Hunter and M.~Taylor,
Phys. Rev. D \textbf{59}, 064005 (1999).

\bibitem{Henn:1985}
  M.~Henneaux and C.~Teitelboim,
  Commun.\ Math.\ Phys.\  {\bf 98}, 391 (1985).

\bibitem{Cald:1999}
M.~M.~Caldarelli, G.~Cognola and D.~Klemm,
Class. Quant. Grav. \textbf{17}, 399
(2000).

\bibitem{Gibb:2004b}
G.~W.~Gibbons, M.~J.~Perry and C.~N.~Pope,
Class. Quant. Grav. \textbf{22}, 1503
(2005).


\bibitem{Suda:1992}
D.~Sudarsky and R.~M.~Wald,
Phys. Rev. D \textbf{46}, 1453
(1992).

\bibitem{Kiri:2009}
 E.~Kiritsis and G.~Kofinas,
  Nucl.\ Phys.\  B {\bf 821} (2009) 467.

\bibitem{Lin:2014}
K.~Lin \textit{et al.},
Int. J. Mod. Phys. D \textbf{23},
1443004 (2014).

\bibitem{Rest:2019}
A.~Restuccia and F.~Tello-Ortiz,
Eur. Phys. J. C \textbf{80},
86 (2020).

\bibitem{Hawk:1971}
S.~W.~Hawking,
Phys. Rev. Lett. \textbf{26}, 1344
(1971).

\bibitem{Penr:1969} R. Penrose, Riv. Nuovo Cimento {\bf 1},
252 (1969).

\bibitem{LIGO:2017}
B.~P.~Abbott \textit{et al.},
Astrophys. J. Lett. \textbf{848},
L13 (2017).

\bibitem{Emir:2017}
A.~E.
G\"umr\"uk\c{c}\"uo\u{g}lu, M.~Saravani and T.~P.~Sotiriou,
Phys. Rev. D \textbf{97},
024032 (2018).

\bibitem{Gong:2018}
Y.~Gong \textit{et al.},
Phys. Rev. D \textbf{98},
104017 (2018).

\bibitem{Will:2014}
C.~M.~Will,
Living Rev. Rel. \textbf{17}, 4 (2014).

\bibitem{Empa:2003}
R.~Emparan and R.~C.~Myers,
JHEP \textbf{09}, 025 (2003).

\bibitem{Fran:2023}
E.~Franzin, S.~Liberati and J.~Mazza,
arXiv:2312.06891 [gr-qc].

\bibitem{Adam:2021}
A.~Adam \textit{et al.},
Class. Quant. Grav. \textbf{39},
125001 (2022).

\bibitem{Park:1998}
M.~I.~Park,
Phys. Lett. B \textbf{440}, 275
(1998);
A.~Strominger,
JHEP \textbf{10}, 034 (2001);
S.~Nojiri and S.~D.~Odintsov,
Phys. Lett. B \textbf{519}, 145
(2001);
D.~Klemm,
Nucl. Phys. B \textbf{625}, 295
(2002);
V.~Balasubramanian, J.~de Boer and D.~Minic,
Phys. Rev. D \textbf{65}, 123508 (2002).

\bibitem{Deru:2004}
N.~Deruelle and J.~Katz,
Class. Quant. Grav. \textbf{22}, 421
(2005);
B.~P.~Dolan,
Class. Quant. Grav. \textbf{36},
077001 (2019).

\bibitem{Aran:2015}
B.~Araneda and G.~Dotti,
Class. Quant. Grav. \textbf{32},
195013 (2015).

\bibitem{Owen:2021}
C.~B.~Owen, N.~Yunes and H.~Witek,
Phys. Rev. D \textbf{103}, 
124057 (2021).




\end{thebibliography}
\end{document}